# Tubulin dipole moment, dielectric constant and quantum behavior:

## computer simulations, experimental results and suggestions.

A.Mershin[1]*,  A.A. Kolomenski[1], H.A. Schuessler[1] and D.V. Nanopoulos[1,2,3]

[1] *Dept. of Physics, Texas A&M University, College Station TX 77843-4242*
[2] *Astro Particle Physics Group, Houston Advanced Research Center (HARC),*
*The Mitchell Campus, Woodlands, TX 77381, USA*
[3] *Academy of Athens, Chair of Theoretical Physics, Division of Natural Sciences,*
*28 Panepistimiou Avenue, Athens 10679, Greece*

## Abstract

We used computer simulation to calculate the electric dipole moments of the $\alpha$ and $\beta$ tubulin monomers and dimer and found those to be $|\boldsymbol{p}_\alpha|$=552D, $|\boldsymbol{p}_\beta|$=1193D and $|\boldsymbol{p}_{\alpha\beta}|$=1740D respectively. Independent surface plasmon resonance (SPR) and refractometry measurements of the high-frequency dielectric constant and polarizability strongly corroborated our previous SPR-derived results giving $\Delta n/\Delta c$ ~1.800x10$^{-3}$ ml/mg. The refractive index of tubulin was measured to be $n_{tub}$ ~2.90 and the high frequency tubulin dielectric constant $\kappa_{tub}$ ~8.41 while the high-frequency polarizability was found to be $\alpha_{tub}$ ~ 2.1x 10$^{-33}$ C m$^2$/V. Methods for the experimental determination of the low-frequency $\boldsymbol{p}$ are explored as well as ways to test the often conjectured quantum coherence and entanglement properties of tubulin. Biobits, bioqubits and other applications to bioelectronics are discussed.

KEYWORDS: Tubulin, electric dipole moment, biobit, bioqubit, quantum, entanglement, surface plasmon resonance, refractometry, refractive index.

*To whom correspondence should be addressed: mershin@physics.tamu.edu

# 1.    Introduction

Although the structure of tubulin, the building block of microtubules (MTs) has been solved by electron crystallography to better than 3.7Å, [1, 2], its electric dipole moment has only been calculated via computer simulations including the ones presented here and in [3].

Knowledge of the tubulin electric dipole moment will be useful in studies of MT (de)polymerization dynamics, simulations of MTs as cellular automata and the experimental testing of hypotheses suggesting quantum behavior, as it can be incorporated in the various models as an experimentally determined parameter. Furthermore, it may help in understanding whether and how to use tubulin self-assembly in novel biomaterial nanofabication applications.

This communication is divided into eight sections. Section 1 gives a short overview of the field. Section 2 presents our computer simulation results for the electric dipole moment and other geometric and energetic properties of the tubulin dimer and monomers. In Section 3, we report on our surface plasmon resonance (SPR) and refractometry measurements of the high-frequency dielectric constant and polarizability of tubulin in solution. Section 4 describes several possible avenues to the experimental determination of the low-frequency dipole moment. Section 5 suggests experimental tests of the often-conjectured ability of tubulin dimers to sustain coherent and/or entangled 'dipole quanta' states [4]. Section 6  discusses the implications for biomolecular electronics while sections 7 and 8 provide a summary of obtained results and conclusions.

## 1.1    Significance and Possible Roles of Tubulin Dipole Moment

There have been some preliminary experiments showing MTs are birefringent [5] and others aimed at measuring the electric field around MTs [6,7,8] indicating that MTs could be ferroelectric (i.e. spontaneous abrupt orientation of tubulin dipoles occurs for an above-threshold externally applied electric field). Some work on the optical properties of MTs appeared in [9,10]. Work by members of this group has concentrated on investigating the effects of MT disruption (by expressing vertebrate microtubule-associated protein (MAP)-tau in mushroom body neurons of –invertebrate- *Drosophila* fruitflies) on the associative memory of these animals. We have concluded that even a small disruption in the microtubular network of neurons causes animals to lose their ability to form and retain associative memories [11] thus supporting hypotheses of MT involvement in memory function. Apart from the above experimental observations, there exists little experimental evidence concerning the electrical or often mentioned quantum properties of tubulin and MTs and how these affect/effect their function. On the other hand, there is a large amount of theoretical work describing various electrical, optical and quantum properties that tubulin and MTs are expected to have based upon their structure and function. Ferroelectricity has been supported by the analysis in [12]. Drug interactions with tubulin have been under investigation and it has been theorized that electric dipole moment 'flips' are responsible for attractive London forces during tubulin binding to other molecules also possessing dipole moments such as general anesthetics [13]. There is controversy as to the correct mechanism of polymerization with the "GTP cap" theory [14] facing alternatives such as those described in [15] and the size of the minimum nucleus required to start polymerization is not clearly understood. The MT paracrystalline geometrical structure has been implicated in error-correcting codes [16]. Intracellular loss-free energy transport along MTs via 'solitonic kink like excitations' or 'dipole quanta' has been theoretically described in [4,17]. In addition to the above mentioned wave phenomena (collectivelly such waves of tubulin dipole moment flips will be called 'flip waves') *all* of the microtubule-based 'quantum brain' models that abound today [4,18,19,20,21,22] have at their core the assumption that tubulin is capable of some sort of conformational changes while in the polymerized state. Note that this applies even to the more exotic models of [19,20] that involve quantum effects arising from the difference in the gravitational states of two tubulin conformations, since the geometrical shifts between the monomers are unlikely to leave the dipole moment vector unchanged. Starting from such assumptions, predictions have been developed such as the existence of long-lived superposed and entangled states among tubulin dimers and long-range non-neurotransmitter based communication among neurons [16,17,18,21]. Further, interaction between water molecule dipoles and the tubulin dipole plays a central role in models predicting emission of coherent photons from MTs[22], intracellular quantum teleportation of dipole quanta states [18] and other controversial yet fascinating features[21].

Today's conventional silicon based devices are of order 180nm in size while future molecular devices promise a further order of magnitude reduction to this minimum. As a result, there have been considerable efforts concentrated on identifying various chemical substances with appropriate characteristics to act as binary switches and logic gates. For instance, rotaxanes have been considered as switches/fuses [23] and carbon nanotubes as active channels in field effect transistors [24]. Many of these substances are unsuitable for placement on traditional chips [25] or for forming networks, while virtually all of these efforts attempt to hybridize some kind of electrical wires to chemical substrates in order to obtain current flows. This complicates the task because of the need for appropriately nanomanufactured wires and connections.

Here we propose an alternative approach. Instead of the binary states being determined by the presence or bulk movement of electrons, they can be defined in terms of the naturally occurring conformational (and consequently dipole moment) states of tubulin or similar protein molecules and their self-assembled polymers. Moreover, the external interaction with these states can be performed by coupling laser light to specific spots of a polymer network. In the case of MTs, signal propagation may be achieved via travelling electric dipole moment 'flip waves' along MTs while modulation may be performed by MAP binding that creates "nodes" in the MT network. In this proposed scheme for information manipulation, there is no bulk transfer of (charged) mass involved. Tubulin polymerization can be controlled by temperature and application of chemicals and MAPs to yield closely or widely spaced MTs, centers, sheets, rings and other structures [26,27], thus facilitating fabrication of nanowires, nodes and networks and even structures capable of long-term information storage (biomolecular computer memory). Since MTs are polar, algnment can be achieved by application of dc electric fields.

Many of these theoretical studies and proposals will be facilitated by a determination of the electric dipole moment of tubulin and its dynamics as this property features prominently in the underlying models.

## 1.2    Properties of Tubulin and Microtubules

Under normal physiological conditions, tubulin is found as a heterodimer, consisting of two nearly identical monomers: α- and β- tubulin, each of molecular weight of about 55kDalton [2]. Electron crystallography measurements on zinc-induced tubulin protofilament sheets [1] have shown that the tubulin heterodimer has dimensions 46x80x65 Å [Fig. 1 (b)] and a 3.7 Å resolution atomic map has been made available in the public [28]. The β- tubulin monomer can bind guanosine 5' triphosphate (GTP), in which case it exists in an energy-rich form that favors polymerization (self assembly), or it can bind guanosine 5' diphosphate (GDP-tubulin) thus being in an energy-poor form (GDP-tubulin) that favors dissociation.

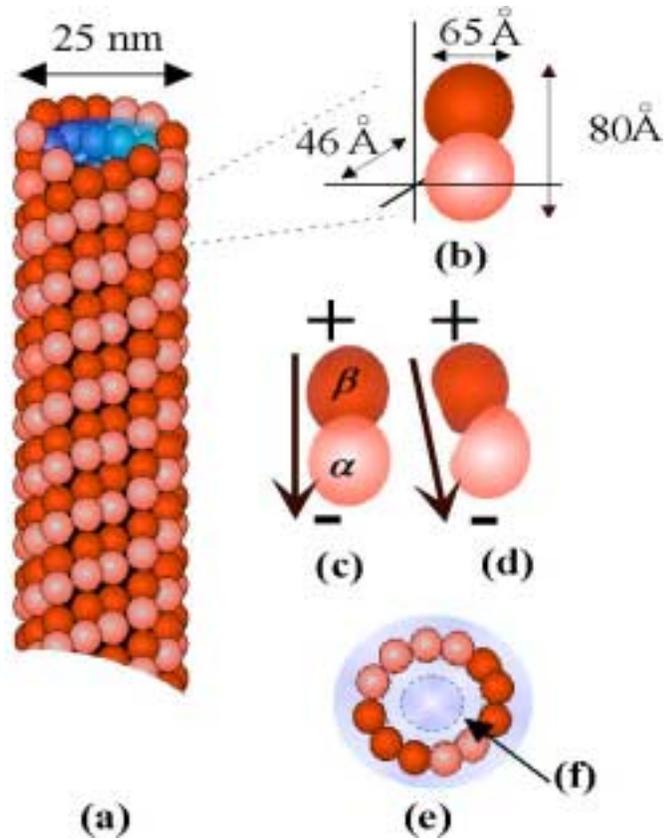

**Figure 1: Tubulin and Microtubule.** Tubulin is a common polar protein found mainly in the cytoskeleton of eukariotic cells and especially enriched in brain and neural tissue. MTs are hollow (25nm-outer diameter, 14nm inner diameter), tubes forming the main component of the cytoskeleton and apart from giving shape and support to the cell, they also play a major role in cellular transport, have been hypothesized to be central in cellular information processing [41] and suggested as possible candidates for biological and/or quantum computation. **(a)** A typical neuronal microtubule made of 13 tubulin protofilaments **(b)** dimensions of the aβ-heterodimer **(c)** GTP-tubulin and **(d)** GDP-tubulin **(e)** a cross section of the MT showing water environment and C-termini extending roughly 6nm away from MT surface note that only 12 (not 13) are shown as this is a cross section **(f)** thin isolated region that we have been theoretically shown to be equivalent to a quantum optical (QED) cavity[4].

Certain interesting phenomena arise during the (de-)polymerization of tubulin such as length oscillations, treadmilling etc. generally referred to as 'dynamic instability'. These effects have been studied extensively [29,30] but are not directly relevant to our analysis at this stage as such phenomena can be avoided during *in vitro* experiments by choosing appropriate buffer environments e.g. adding chemicals such as taxol or glycerol that stabilize microtubules in the polymerized state. The GDP-GTP exchange (hydrolysis) releases ~0.4eV per molecule and is accompanied by a conformational change [10]. This change has been modeled as resulting in a ~27$^o$ angle [31] between the original line connecting the centers of the α- and β- monomers and the new center-to-center line [Fig. 1 (d)].

MTs come in a variety of arrangements the predominant of which is a 5-start, period-13 helical tube of dimers which resembles a corn ear [Fig.1 (a)] made out of 13 offset protofilaments. At physiological pH (=7.2) MTs are negatively charged [32] due to the presence of a 15-residue carboxyl terminus 'tail' and at pH=5.6 MTs become neutral. There have been suggestions that the C-terminus [see Fig. 1 (e)] is important in polymerization, protein interactions and perhaps charge conduction [33]. This terminus has not been included in the electron crystallography data of [2] so all values concerning the dipole moment vector **p** are quoted with the understanding that **p** has been calculated ignoring the effect of the C-terminus. This necessary approximation is partially warranted because of the symmetrical nature of the C-terminus tails extending outward from the MT trunk and conceivably canceling each other's electrostatic contribution at least at the MT interior [see Fig.1 (f)].

## 2. Computer Simulation

A naïve estimate of the tubulin dipole moment **p** in vacuum based on a charge of 10 electrons multiplied by a separation of 4nm (roughly the distance between the monomers) yields an |**p**| of 1920D. Similarly, a rough volume estimate is V = 2$V_{glob}$ where $V_{glob}$ refers to the approximately spherical volume of each globular monomer of radius R≈26 Å so that $V_{glob}$=$^4/_3\pi R^3$ ≈73,600 Å$^3$. The TINKER V3.9 (2001) molecular modeling package [34,35] was applied with the CHARMM set of force-field parameters [36] to the 1TUB protein database data (pdb) [28] consisting of some 17,000 atomic coordinates and we calculated |**p**| to be 1740D for the tubulin dimer in close agreement with previous molecular simulation calculations [36].

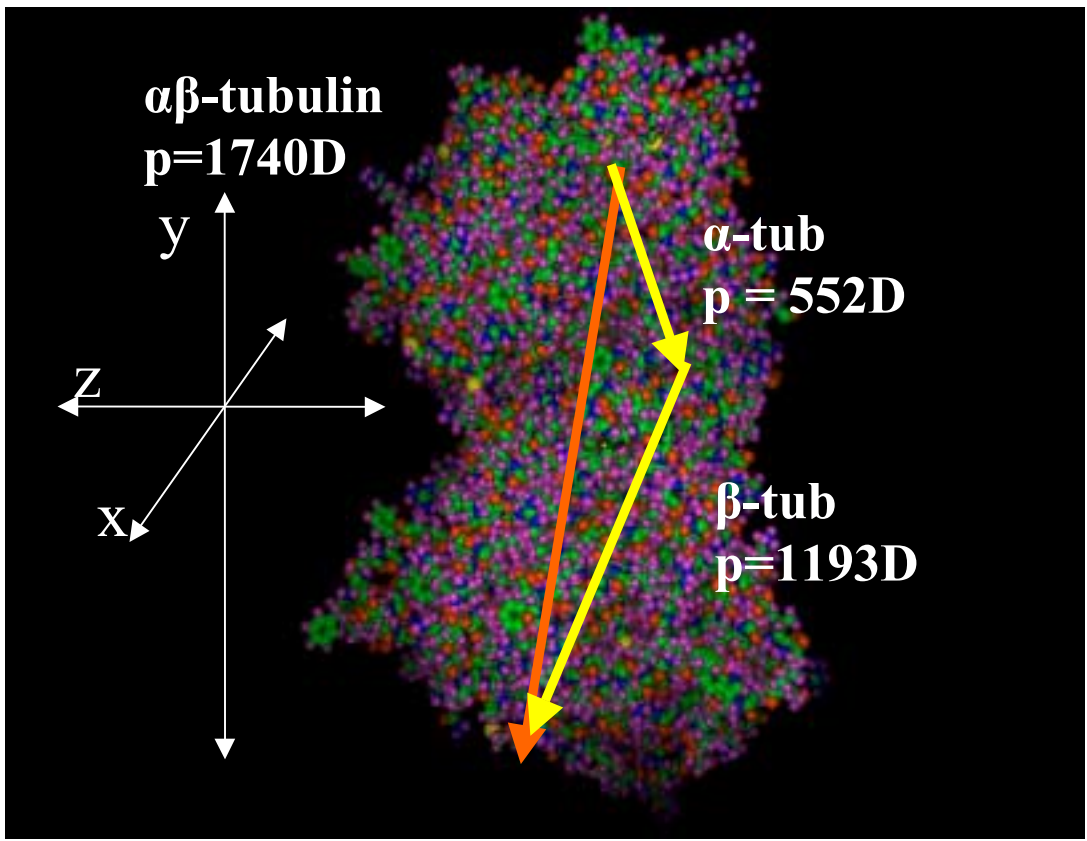

**Figure 2. A schematic of the roughly 17,000 atoms of tubulin and the electric dipole moments of the dimer and monomers.**

(Polar angles with MT aligned along y-axis)

**αβ-Tubulin Dimer**

| | | |
|---|---|---|
| Total Electric Charge | : | -10e |
| Electric Dipole Moment Magnitude | : | 1739 Debyes |
| Electric Dipole Moment Direction | : | $\theta = 83.02^o$ ; $\varphi = 82.97^o$ |
| Radius of Gyration | : | 28.6 Å |

**α-Tubulin Monomer**

| | | |
|---|---|---|
| Total Electric Charge | : | -5.0e |
| Electric Dipole Moment Magnitude | : | 552 Debyes |
| Electric Dipole Moment Direction | : | $\theta = 79.02^o$; $\varphi = 75.21^o$ |
| Radius of Gyration | : | 20.6 Å |

**β-tubulin monomer**

| | | |
|---|---|---|
| Total Electric Charge | : | -5.0 |
| Electric Dipole Moment Magnitude | : | 1194 Debyes |
| Electric Dipole Moment Direction | : | $\theta^o = 79.7^o$; $\varphi = 75.21^o$ |
| Radius of Gyration | : | 20.6 Å |

TINKER with CHARMM parameters has given close matches to experimental values for dipole moments of other molecules [3], so we expect the above quoted $|\mathbf{p}_{\alpha\beta}|$=1740D to be reasonably close to the real value of the dipole for the crystallized tubulin unit cell in a flat sheet of antiparallel zinc-induced protofilaments, without C-termini, (as these were the experimental conditions for the pdb data). It was also found that the dipole moment of the individual α- and β-monomers to be $|\mathbf{p}_{\alpha}|$=552D and $|\mathbf{p}_{\beta}|$=1193D respectively. Note that the β-monomer has approximately twice the dipole moment vector magnitude of the α-monomer and both point in virtually the same direction, consistent with the fact that they have nearly identical sequence and tertiary structure. The minimum volume we calculated using pdb data was V≈78,000 $\text{Å}^3$. As expected, the Van der Waals attractive forces constitute the maximum contribution to the energy of the tubulin molecule.

## 3.    Experimental Methods and Results

### 3.1    High frequency tubulin dielectric constant and polarizability determination.

The dipole moment is intimately related to the dielectric constant $\kappa$ through:

$$\boldsymbol{P} = (\kappa\text{-}1)\ \varepsilon_o \boldsymbol{E}$$

where $\boldsymbol{P}$ is the total polarization of a sample, $\kappa = \varepsilon\ /\varepsilon_o$ (where $\varepsilon$ is the permittivity of the material and $\varepsilon_o$ the permittivity of vacuum) and $\boldsymbol{E}$ is the local electric field vector. Therefore as a preliminary step to the eventual electric dipole moment determination, we reported in [38] on our SPR-based measurement of the slope of the refractive index $n$ versus concentration $c$ straight line for tubulin:

$$\frac{\Delta n}{\Delta c} = (1.85 \pm 0.20) \times 10^{-3} \left( mg\ /\ ml \right)^{-1} \Rightarrow \frac{\Delta \varepsilon}{\Delta c} = (5.0 \pm 0.5) \times 10^{-3} \left( mg\ /\ ml \right)^{-1} \quad (1)$$

Note that for optical frequencies, the dielectric constant is related to the refractive index n via

$$\kappa = n^2 \qquad\qquad\qquad\qquad (2)$$

Here we corroborate this result with a refractometry measurement as follows. We used a commercial refractometer [Abbe Refractometer, Vista C10] of exceptional accuracy. Briefly, light was allowed to enter and be reflected into a prism, which was coated with the sample of interest and covered. The prism's refractive index was known and the beam's incident angle was tilted until total internal reflection was reached (seen as a dark band in the eyepiece). The refractive index of the sample was then read on a pre-calibrated scale. This method depends on the prism having a higher refractive index than the sample. After standard calibration and prior to measurement of tubulin solutions, a number of NaCl solutions of varying concentration were used as additional calibration (data not shown). Determining the exact concentration was the main source of error in this measurement so high precision electronic scales and precision micro pipettes were used. The prism was cleaned after each measurement with ethanol soaked cotton and left to dry before applying the next sample. It was found that 30-50μL of solution were adequate to deposit a thin film on the prism such that there was virtually no noise (indicated as colors). This small volume is comparable to the requirements of the sophisticated SPR-based BIAcore 3000 [39] machine for a single injection.

The refractive indices of a series of concentrations of NaCl (in 18.1MΩ $H_2O$) and tubulin in the same buffer as for the SPR measurement (0.1M 4-Morpholinoethane sulphonic acid, 1mM EGTA, 0.1mM EDTA, 0.5mM$MgCl_2$, 1mM GTP at pH 6.4) were measured. Three measurements were taken for each data point. Errors are estimated at ~5% for concentration (see Fig. 3). The precision of the refractometer was scale limited with an error of 0.00005 for n (not shown, represented as the size of the data points). In Figure 3 a least-squares fit linear regression yields straight lines with an R factor of 0.9928. The intercepts were manually set to the zero-point concentration averages. The results at different times (i.e. pre- and post-polymerization) did not deviate appreciably suggesting that at this wavelength range, solutions of tubulin dimers and microtubules have similar refractive indices.

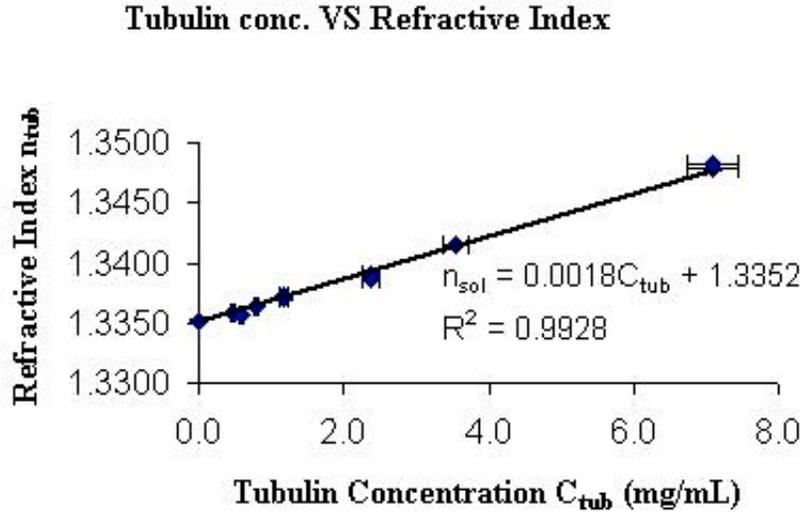

**Figure 3. Refractive Index VS Tubulin Concentration.**

A very small change in concentration of tubulin results in a significant change in the index of refraction (Fig. 3) giving a slope of $\Delta n/\Delta C = 1.800 \pm 0.090 \times 10^{-3}$ ml/mg (this result is strongly corroborated by our previous SPR measurement -equation (1): $\Delta n/\Delta C$ $1.80 \pm 0.20 \times 10^{-3}$).

A physiological concentration is assumed to be 15.0μM (i.e. $1.50 \times 10^{-5}$ mol/L) [40]. Since the molecular weight of a tubulin dimer is 110kD this gives a proportionality of 1.00mg/mL $\sim$ 9.10 μM i.e. 15.0μM is equivalent to 1.60mg/mL. This results in a molecular density of $N = 9.03 \times 10^{18}$ tubulin molecules per liter or $9.03 \times 10^{21}$ m$^{-3}$, but the concentration necessarily varies across cell types, intracellular position and cell condition. For instance, although a MAP-tau-burdened cell has the same overall tubulin density, it has much higher local axonal and dendritic density of tubulin when MTs have deteriorated into neurofibrillary tangles as is the case in Alzheimer's Disease.

The partial contribution to the refractive index of the solution by tubulin can be found if the density dependence on concentration of the solution is known (see Fig. 3). Assuming that the contributions from the various components are additive linearly, we have for the total index of refraction of the solution $n_{sol} = \sum_i C_i n_i$ , where $C_i$ and $n_i$ are the fractional concentration of the $i_{th}$

component with refractive index $n_i$ and i runs over all components. Lumping the contribution

from all the buffer components we can write

$$n_{sol} = (1 - \chi_{tub})n_{buffer} + \chi_{tub}n_{tub} \Rightarrow n_{tub} = \frac{n_{sol} - (1 - \chi_{tub})n_{buffer}}{\chi_{tub}} .$$

Where $\chi_{tub} = C_{tub}/\rho$ the mass fraction and $\rho$ is the density of the solution. It may be argued that using a volume fraction is more appropriate here but for our purposes a mass fraction is adequate and simpler. At $C_{tub} = 1.60$ mg/mL, $n_{sol} = 1.8000 \times 10^{-3} C_{tub} + 1.3352$ (from Fig. 3) gives: $n_{sol} = 1.34$

$\pm 0.07$ and using solution density $\rho = 1.45$ gr/mL[41], at $C_{tub} = 1.60$ mg/mL we arrive at the value for

$n_{tub}$:

$$n_{tub} = 2.90 \pm 0.10 , \qquad\qquad (3)$$

which can be used in $\kappa = n^2$ to give the high frequency tubulin dielectric constant:

$$\kappa_{tub} = 8.41 \pm 0.20 . \qquad\qquad (4)$$

To our knowledge, this is the first time these two quantities have been experimentally determined for tubulin.

Both $n$ and $\kappa$ are at the very top range of what is usually assumed for proteins, as expected since tubulin seems to have such a high dipole moment in molecular dynamics simulations.

The refractive index $n$ of an optically dense material is related to the high frequency

polarizability $\alpha$ via:

$$n^2 = 1 + \frac{N\alpha}{\varepsilon_o} \frac{1}{\left(1 - \frac{N\alpha}{3\varepsilon_o}\right)} ,$$

where $N$ is the molecular concentration in which for our chosen concentration is $9.03 \times 10^{21}$

molecules/m$^3$. Solving the above equation for $\alpha$ gives

$$\alpha = \frac{\varepsilon_o}{N} \frac{3(n^2 - 1)}{(n^2 + 2)}$$

and therefore the high frequency tubulin polarizability is

$$\alpha_{tub} = 2.1 \pm 0.1 \times 10^{-33} \text{ C m}^2/\text{V} . \qquad (5)$$

A very large number owing to the evidently large dipole moment of tubulin.

A determination of the high-frequency electric dipole moment $\boldsymbol{p}$ from $\alpha$ is now possible through the approximation $\boldsymbol{p} = \boldsymbol{p}_{perm} + \alpha \boldsymbol{E}_{loc}$ where $\boldsymbol{E}_{loc}$ is the local incident field and $\boldsymbol{p}_{perm}$ is the permanent or low-frequency "dc-" dipole moment that was earlier calculated by simulation to be of the order of 1700D. Next we will suggest several ways of experimentally determining this low-frequency or dc- $\boldsymbol{p}$.

**4.     Suggestions for the experimental determination of  the low-frequency electric dipole moment of tubulin**

**4.1     In vacuum**

Although the most direct approach to determine the electric dipole moment of a protein such as tubulin would be to measure the acceleration of evaporated single molecules in the gradient of an electric field in vacuum, this proves to be a difficult task. Even though it is possible to keep tubulin from polymerizing in solution (e.g. by lowering the temperature or raising the salinity), evaporating individual tubulin molecules is very difficult due to tubulin's affinity towards polymerization and aggregation. In other words, the protein is naturally 'sticky' and will be hard to corpusculize. In addition, any spraying of this large molecule is bound to break it into its constituent peptides.  Furthermore the tubulin electric dipole moment in vacuum is not directly relevant to its physiological value, since ionic influences in solution significantly influence the charge distribution.

**4.2     DC Measurement of ε in Dilute Solution**

In principle, the capacitance $C$ of a flow cell with conducting parallel-plate walls can be measured first filled with air and then filled with the tubulin-buffer solution for various tubulin concentrations. Resistance $R_1$ is assumed to be infinite in the case of air and some finite but very large value (of the order of 10MΩ) in the case of solution. $R_2$ is set to a well known value (e.g. 5kΩ) and is needed to overwhelm any small conductance due to the presence of liquid between the plates so that the equivalent resistance in the circuit is identical for all the measurements. An inductance-resistance-capacitance (LRC) bridge can be used to perform the measurement at several low to medium frequencies (e.g. 100Hz to 1kHz) by measuring the $RC$ time constant and the total equivalent resistance and capacitance for the circuit. The ratio $C'/C$ for the tubulin/air capacitances then gives $\kappa$ and from equation (4) the dipole moment can be inferred. However, we have determined that this simple $RC$ circuit with low inductance connected to a bridge is not ideal because aqueous solutions tend to form time-dependent polarization layers on the electrodes complicating interpretation of the frequency dependence of results [42]. A refined version of this experiment based on our ongoing efforts at dielectric spectroscopy measurements will be reported elsewhere.

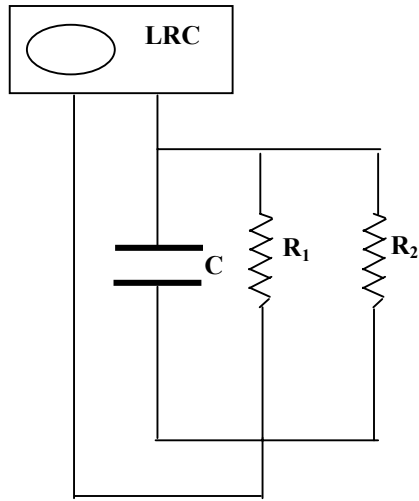

**Figure 4.LRC Bridge for Capacitance Measurement**

# 5. Suggestions for the Experimental Determination of Possible Quantum Effects in Tubulin

## 5.1 Tubulin and Quantum Coherence

In [4], a comprehensive model conjecture treating certain regions inside MTs as isolated high-Q(uality) QED cavities was put forth as well as a scenario according to which the presence of ordered water in the interior of MTs results in the appearance of electric dipole quantum coherent modes, which couple to the unpaired electrons of the MT dimers via Rabi vacuum field couplings. The situation is analogous to the physics of Rydberg atoms in electromagnetic cavities [43]. In quantum optics, such couplings may be considered as experimental proof of the quantized nature of the electromagnetic radiation. In our case, therefore, if experimentally detected, such couplings would indicate the existence of coherent quantum modes of electric dipole quanta in the ordered water environment of MT, as conjectured in [44,45], and used here. To experimentally verify such a situation, one must first try to detect the emergent ferroelectric properties of MTs, which are predicted by this model and are potentially observable and for this measurement of the dipole moment of the tubulin dimer is an important step. A suggestion along these lines has been put forward in [12]. In addition, one should verify the aforementioned vacuum field Rabi coupling (VFRS), $\lambda_{MT}$, between the MT dimers and the ordered water quantum coherent modes. The existence of this coupling could be tested experimentally by the same methods used to measure VFRS in atomic physics [46], i.e. by using the MTs themselves as cavity environments and considering tunable probes to excite the coupled dimer-water system. Such probes could be pulses of (monochromatic) light coupling to MTs. This would be the analogue of an external field in the atomic experiments mentioned above. The field would then resonate, not at the bare frequencies of the coherent dipole quanta or dimers, but at the Rabi splitted ones, leading to a double peak in the absorption spectra of the dimers [46]. By using MTs of different sizes one could thus check on the characteristic √N-enhancement of the (resonant) Rabi coupling: $\Omega = \omega_o \pm \lambda\sqrt{N}$ as described in [4] for MT systems with N dimers.

## 5.2    Tubulin and Quantum Entanglement

The obvious objection to suggestions of long decoherence times for quantum properties of large molecules at room temperature comes from the application of equilibrium principles to the quantum mechanical aspects of the constituent atoms [47]. A ground-breaking experiment has recently been performed and showed the existence of long-lived (t ~ 0.5ms) partially entangled states of the bulk spin of two macroscopically separated samples of $10^{12}$ Cesium atoms at room temperature [51]. We hope to investigate deeper, as although the tubulin molecule consists of some 17,000 atoms which are subject to considerable thermal noise, the electric dipole moment state depends crucially on only a few electrons that can be in two sets of orbitals. In addition, tubulin *in vivo* is not an equilibrium system, rather it is a dynamic dissipative system where energy is being pumped in and out constantly. Our theoretical work has suggested that for a certain set of parameters (such as the value of the dipole moment, the pH etc.) tubulin could indeed sustain a quantum mechanically coherent state for times of the order of microseconds [4].

Since 1935 when Erwin Schroedinger coined the word "entanglement" to refer to a state where the wavefunction describing a system is unfactorizable, much has been learned about this peculiar phenomenon and it has turned out to be very useful in quantum information science, quantum cryptography and quantum teleportation. Entanglement has been experimentally realized in light [48,49], in matter [50] and in combinations of those [51,52]. One way to produce entangled states in light is via type II phase-matching parametric downconversion which is a process occurring when ultraviolet (UV) laser light is incident on a non-linear beta-barium borate (BBO) crystal at specific angles. A UV photon incident on a BBO crystal can, sometimes, due to a non-linear parametric process, spontaneously split into two correlated infrared (IR) photons (each of half the energy of the incident photon). The infrared photons are then emitted on opposite sides of the UV pump beam, along two cones, one of which is horizontally polarized and the other vertically. The photon pairs that are emitted along the intersections of the two cones have their polarization states entangled. This results in each photon being arbitrarily polarized, but the two photons belonging to a pair necessarily have perpendicular polarizations to each other. The state $\Psi$ of the outgoing entangled photons can be written as: $\left| \Psi \right\rangle = \left( \leftrightarrow, \updownarrow \right) + e^{i\alpha} \left( \updownarrow, \leftrightarrow \right)$ where the arrows indicate polarizations for the (first, second) IR photon and can be controlled by inserting appropriate half wave plates, while the phase factor $e^{i\alpha}$ can be controlled by tilting the crystal or using an additional BBO in a setup similar to the one depicted in Fig. 5, modified from [53]

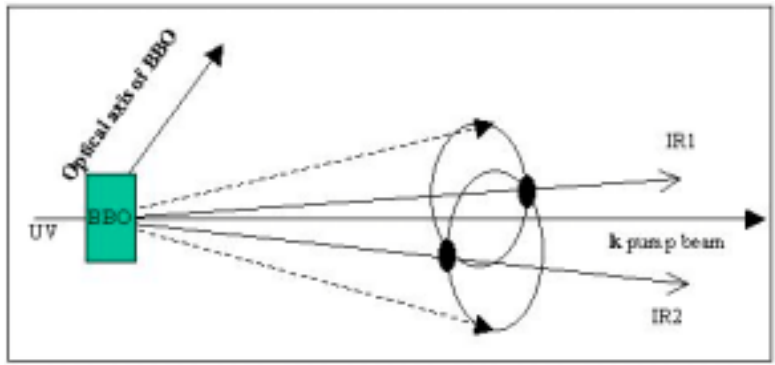

**Figure 5. Type II Phase Matching Parametric Downconversion.** For certain orientations, a UV photon is converted by the non-linear BBO crystal into two entangled IR photons (IR1, IR2)

Measuring the state of one of the outgoing photons -say IR1, immediately determines the state of the other (IR2) regardless of their separation in space. This counterintuitive phenomenon is referred to as the Einstein Podolsky Rosen (EPR)-paradox and such pairs are called EPR pairs.

An on-demand entangled photon source (ODEPS) can be built and utilized to do spectroscopic analysis of proteins and other biomolecules. Such a device would greatly facilitate studies of the fundamental quantum properties of entangled objects, possible quantum properties of living matter, quantum information science and more. It would also provide an opportunity to test and identify the problems associated with the construction of a portable entanglement source to be taken "into the field" if and when entangled photons become indispensable to secure communications. For instance, tolerances of entangled photon generation to vibration and temperature changes could be studied.

## 5.3    ODEPS/SPR Determination of Conjectured Quantum Nature of Tubulin.

A recent experiment titled "plasmon-assisted transmission of entangled photons" [54] has shown that originally entangled photons stay entangled after being converted to (entangled) surface plasmons and back, at room temperature.

A minimally modified design of an SPR biosensor apparatus such as the one used in our measurement of $\varDelta n/\varDelta c$ in [38] coupled to an ODEPS described in section 5.2 above, could be capable of detecting the often conjectured mesoscopic bulk coherence and partial quantum entanglement of electric dipole moment states, existence of which would cast biomolecules as serious candidates for the implementation of qubits.

The experiment can be performed as follows. Under a protocol similar to one developed by Oberparleiter et al. [55], a source producing in excess of 360,000 entangled photon pairs per second, can be coupled to a setup similar to the one developed by Altewischer et al. [54] where entangled photons are transduced into (entangled) surface plasmons and re-radiated back as entangled photons. The essential difference would be that the insides of the perforations in the gold film of Altewischer et al. would be covered with a monolayer of tubulin dimers or microtubules immobilized by standard methods similar to the ones described in [38]. The evanescent wave of the (entangled) surface plasmon generated at resonance will interact with the electric dipole moment of the immobilized protein complexes and presumably transfer the entanglement to a dipole state in a manner similar to the transfer of the photon polarization entanglement to surface plasmons.

At the other end of the perforations, the surface plasmons would be reradiated as photons having undergone the interaction with the protein electric dipole moment. If (partial) entanglement with the partner photon (that underwent none of these transductions) is found, then this would suggest that the protein is capable of partially "storing" the entanglement in its electric dipole moment state and characteristic decoherence times could be determined. Note that this measurement is capable of directly testing the hypothesis of [13] that general anesthetics act by shutting down such electrical and entanglement activities in MTs since application of anesthetic gas should presumably destroy any entanglement survival.

## 6.      Discussion

. It is conceivable that the role of the ubiquitous binary state necessary for computation can be decoupled from the presence or bulk movement of charge if one can assign it to the naturally-occurring conformational states of protein molecules. This scheme would be qualitatively similar to efforts in the new field of spintronics where the spin-polarization of electrons inside modified semiconductor-based logic gates is the defining quantity for electron dynamics and switching [56].

Regardless of whether it turns out that tubulin and MTs are purely classical systems or they have a quantum nature, the excitation and detection of the theory-suggested 'flip waves' [18] would be an important step towards understanding the role that tubulin and MTs can play as binary switches and networks respectively, both in naturally occurring systems such as living cells as well as in synthesized structures. Note that the energy needed for a tubulin conformational change or 'flip' is roughly 200 times lower than a conventional silicon-based binary switch, making laser-pulse induced switching feasible. This conformational change energy is also about 30 times larger than thermal noise at room temperature, making the system reasonably resilient to thermal noise. Certain structures other than straight MTs, for instance single and double rings, sheets, spirals, centers etc. can also be induced to self-assemble by using well-understood biochemical protocols and we expect these to eventually play distinct roles in building microelectronic components. Note that although biochemically it seems MTs are made predominantly of GDP-tubulin, this does not mean that the dipole flips are 'frozen-out' since intramolecular electron motion can still occur in the hydrophobic pockets of the tubulin heterodimer without the need for nucleotide binding

The techniques described here would also be suited to check several predictions of recent theories regarding quantum effects in the cytoskeleton. The existence of 'flip waves' along filaments of tubulin molecules is an important assumption of many current theories of MT-network function as a computational and information-processing system. To check this hypothesis. The interaction of a longitudinal component of the field of a surface plasmon with a dipole moment may provide a coupling mechanism for the excitation and detection of these waves in future studies. At present, predictions of the phase velocity of flip waves differ drastically. Depending on the model and the parameters assumed, the speed of such waves has been estimated to be between 1 and 800 m/s. The external interaction with these tubulin states could be performed by coupling laser light to specific spots of the MT network. Signal propagation could be achieved by travelling electric dipole moment flip waves along protofilaments and MTs while modulation can be achieved by MAP binding that creates "nodes" in the MT network.

## 7.    Summary

Theoretical efforts by us and others have strongly suggested that tubulin is near the "front lines" of intracellular information manipulation and storage. Our group has performed preliminary measurements on tubulin in an effort to supply experimentally determined parameters (such as the refractive index, polarizability and dipole moment) to apply to the various models of tubulin. In addition, it has become increasingly evident that fabrication of novel biomaterials through molecular self-assembly (e.g. microtubules) is going to play a significant role in material science [57] and possibly the information technology of the future [58]. Here we have suggested that relatively straight-forward spectroscopic techniques such as refractometry, surface plasmon resonance sensing and dielectric spectroscopy, coupled with molecular dynamic simulations and (quantum) electrodynamic analytical theory are useful tools in the study of dielectric and quantum properties of proteins and the first steps in those directions have been illustrated.

We used computer simulation to calculate the electric dipole moments of the two tubulin monomers and dimer and found those to be $|p_\alpha|$=552D, $|p_\beta|$= 1193D and $|p_{\alpha\beta}|$=1740D respectively. We used refractometry to corroborate our previous SPR-derived result (equation (1)) for $\Delta n/\Delta c$ ~1.800ml/mg. The refractive index of tubulin was found to be $n_{tub}$ ~2.90 (3) and that gives the high frequency tubulin dielectric constant at $\kappa_{tub}$ ~8.41 (4). In addition, the high-frequency polarizability was found to be $\alpha_{tub} \sim 2.1 \times 10^{-33}$ C m$^2$/V (5). Several methods were described to determine the low-frequency DC-$p$ as well as to check for both coherence and entanglement among tubulin dimer dipole states. An experiment was suggested whereby using a perforated metal chip layered with a network of aligned MTs, and employing entangled photons in the SPR-exciting laser beam, it can be determined whether surface plasmons interacting with MTs can stay entangled and whether this entanglement can be propagated and conserved by the biomolecules. Possible applications of tubulin and microtubules to the bioelectronics of the future were discussed.

## 8. Conclusion

The electric and energy-transduction properties of tubulin and the polymers it forms are important not only because of the role these play in intracellular protein interactions but also because it may well be that nature has already provided us with suitable nanowires, switches or even logic gates. Beyond the obvious benefit to the credibility or otherwise of the various "quantum brain models", determining the dipole moment of tubulin and its dynamics will further our understanding of tubulin and other similar proteins (such as actin) and will shed light on whether we can use these as the basis of biomolecular electronic circuits of even quantum information processing.

Tubulin, microtubules and the dynamic cytoskeleton are fascinating systems and in their structure and function contain the clues on how to imitate nature in artificially fabricated biomolecular information processing devices paving the way for biobits and perhaps even bioqubits.

## 9. Aknowledgements

This material is based upon work supported by the National Science Foundation (NSF) under Grant No. 0218595 and by a Texas Informatics Task Force (TITF) grant. AM is partially supported by the A.S. Onassis Public Benefit Foundation (Greece). We wish to thank Dr. L. Perez for her assistance with molecular simulation software.

**References**

1.      Nogales, E., Whittaker, M., Milligan, R.A., Downing, K.H., "High Resolution Model of the Microtubule", *Cell* **96**: 79-88 (1999)

2.      Nogales E, Wolf S.G., Downing K.H., "Structure of the αβ tubulin dimer by electron crystallography". *Nature* **291**:199-203 (1998)

3.      Brown, J. A., *PhD Thesis*, University of Alberta (Edmonton) (1999).

4.      Mavromatos N.E. and Nanopoulos D.V. "On Quantum Mechanical Aspects of Microtubules" *Inter.J.of Mod. Physics B* **B12**: 517 (1998)

5.      www.foresight.org/ Conferences/MNT6/Abstracts/Unger/index.html]

6.      Pokorny J., Jelinek F., Trkal V. "Electric field around microtubules" *Bioelectrochemistry and Bioenergetics* **45:** 239-245 (1998)

7.      Jelinek F,  Pokorny J.,  Saroch J., Trkal V. Hasek J., Palan B.,  "Microelectronic sensors for measurement of electromagnetic fields of living cells and experimental results" *Bioelectrochemistry and Bioenergetics* **48:** 261-266 (1999)

8.      Pokorny J. "Conditions for coherent vibrations in the cytoskeleton" *Bioelectrochemistry and Bioenergetics* **48:** 267-271 (1999)

9.      Ventilla, M., Cantor, C., R., & Shelanski, M. "A circular Dichroism Study of Microtubule Protein" *Biochemistry* 11 No. 9 (1972)

10.     Audenaert, R., Heremans, L., Heremans K., & Engelborghs Y. "Secondary structure analysis of tubulin and microtubules with Raman spectroscopy" *Biochimica et Biophysica Acta - Protein Structure and Molecular Enzymology,* 996, Issues 1-2, 13, 110-115 (1989)

11.     Mershin, A., Pavlopoulos, E., Fitch, O., Braden, B.C., Nanopoulos, D.V. & Skoulakis, E.M.C. "Learning and Memory Deficits upon TAU accumulation in Drosophila

Mushroom body Neurons" *Accepted to J. Learning and Memory Oct. 2003 -expeted pub. date 02/2004.*

12. Nanopoulos D.V. Mavromatos N.E.  Zioutas, K. "Ferroelectrics and their possible involvement in biology" *Advances in Structural Biology*, **5**: 127 (1998).

13. Hameroff, S. "Anesthesia, consciousness and hyrdophobic pockets –a unitary quantum hypothesis of anesthetic action", *Toxicology Letters* **100-101**: 31-39 (1998)

14. Mitchison, J, "Microtubule Polymerization Dynamics", *Annu. Rev. Cell Dev. Biol*. **13:** 83-117, esp sections ″The GTP CAP Model″ p95-97, ″Structural Basis of Dynamic Instability″ pp97-99 and ″Relationship of Structural and Chemical Transitions″ pp99-100 (1997)

15. Hyman, A.A., Chretien, D., Arnal, I., Wade, R.H.,  "Structural Changes Accompanying GTP Hydrolysis in Microtubules: Information from a Slowly Hydrolyzable Analogue Guanulul-(a,b)-Methylene-Diphosphonate" *The Journal for Cell Biology* **128** Numbers 1& 2: 117-125 (1995)

16. Mershin, A., Nanopoulos D. V. , Skoulakis E. M. C. Quantum Brain? *Proceedings of the Academy of Athens* **74**, ACT-08/00, CTP-TAMU-18/00 (1999), e-reprint: http://arXiv.org/abs/quant-ph/0007088

17. Sataric, M.V., Tuszynski, J.A., Zakula, R.B. "Kinklike excitations as an energy transfer mechanism in microtubules" *Physical Review E* **48**, No. 1: 589-597 (1993)

18. Mavromatos, N.E. , Mershin, A., & Nanopoulos, D.V. ″QED-cavity model of microtubules implies dissipationless energy transfer and biological quantum teleportation″ International Journal of Modern Physics B, **16**, No. 24 (2002) 3623-3642

19. Penrose, R. (1989) *The Emperor's New Mind,* Oxford University Press, Oxford.

20. Penrose, R. (1994) *Shadows of the Mind* Oxford University Press, Oxford

21.     Hameroff, S., Nip, A., Porter, M., Tuszynski, J.A. *BioSystems* **64**, 149-168 (2002)

22.     Jibu, M., Brinbram, K.H. & Yasue, K.. "From conscious Experience to Memory Storage and Retrieval: The role of quantum brain dynamics and boson condensation of evanescent photons" International Journal of Modern Physics B **40** Nos. 13 & 14 1735-1754 (1996)

23.     Collier, C. P., Wong, E. W., Belohradský, M., Raymo, F. M., Stoddart, J. F., Kuekes, P. J., Williams, R. S., Heath, J. R. "Electronically Configurable Molecular-Based Logic Gates" *Science* 285: 391-394 (1999)

24.     Gates V. Derycke, R. Martel, J. Appenzeller, Ph. Avouris "Carbon Nanotube Inter- and Intramolecular Logic", *NanoLetters* (10.1021/nl015606f S1530-6984(01)05606-5) **Vol. 0**, No.0, p.3.3 (2001)

25.     Kohzuma, T. Dennison, C. McFarlane, W. Nakashima, S. Kitagawa, T. Inoue, T. Kai, Y. Nishio, N. Shidara, S. Suzuki,  S., Sykes A.G., "Spectroscopic and Electrochemical Studies on Active-site Transitions of the Type 1 Copper Protein Pseudoazurin from Achromobacter cycloclastes**"** *J. Biol. Chem*. **270**: 25733-25738.

26.     Diaz, J.F. Pantos, E. Bordas, J., Andreu, M.J. "Solution of GDP-tubulin Double Rings to 3nm Resolution and Comparison with Microtubules" *J. Mol. Biol* **238**: 214-225 (1994)

27.     Hirokawa, N., Shiomura, Y., Okabe, S. "Tau proteins: The Molecular Structure and Mode of Binding on Microtubules" *The Journal of Cell Biology* **107:** 1449-1459 (1988)

28.     Protein Data Bank domain I.D.:1TUB; Nogales et. al. 1998

29.     Jobs, E., Wolf, D.E., and Flyvbjerg, H., "Modeling Microtubule Oscillations", *Phys. Rev. Lett*. **29**, No.3.: 519-522 (1997)

30.     Flyvbjerg, H., Holy, T.E., and Leibler, S., "Stochastic Dynamics of Microtubules: A Model for Caps and Catastrophes", *Phys. Rev. Lett*. **73**, No. 17: 2372 – 2375 (1994)

31. Melki, R., Carlier, M.F., Pantaloni, D., Timasheff, S.N. "Cold Depolymerization of Microtubules to double rings: Geometric Stabilization of Assemblies", *Biochemistry* **28**: 9143-9152 (1989)

32. Stebbins, H., Hunt, C. "The nature of the clear zone around microtubules" *Cell. Tissue Res*. **227**:609-617 (1982)

33. Sackett, D.L., Structure and Function in the "Tubulin Dimer and the Role of the Acidic Carboxyl Terminus" *Subcellular Biochemistry, Vol. 24*. Proteins: Structure, Function and Engineering, Biswas B.B., Roy, S. Eds., Plenum Press, New York (1995)

34. Duffy J.B. (2002) *Genesis* **34,** 1-15.

35. Ponder J.W. & Richards F.M. (1987) *J. Comput. Chem.* **8**, 1016-1024.

36. Pappu, R.V., Hart, R.K., Ponder, J.W., (1998) *J. Phys. Chem. B,* **102,** 9725-9742.

37. A. Mershin website URL      http://people.physics.tamu.edu/mershin

38. Schuessler, H.A. , Mershin, A., Kolomenski A.A. & Nanopoulos, D.V. Surface plasmon resonance study of the actin-myosin sarcomere complex and tubulin dimer J. Mod. Optics, **50** No. 15-17, 2381-2391 (2003)

39. BIAcore, Inc. "BIAcore AB": http://www.biacore.com/scientific/reflist.html (accessed on 06/01/2002).

40. Vater, W., Bohm, K.J., & Unger E., (1997) *Cell Mobility and the Cytoskeleton* **36,** 76-83.

41. Quillin, M.L. & Matthews, B.W. (2000) *Acta Crystallogr D Biol Crystallogr*. **56 ( Pt 7),** 791-794.

42. Lioubimov, V. Kolomenskii, A.A, Mershin, A.,  Nanopoulos, D.V. and Schuessler H., A., The Effect of Varying Electric Potential on Surface Plasmon Resonance Sensing, *(submitted to J. Applied Optics Dec. 2003)*

43. Sanchez-Mondragon, J.J. Narozhny N.B. & Eberly, J.H. (1983) *Phys. Rev. Lett.* **51**, 550-560.

44. Del Giudice, E., Doglia, S., Milani M. & Vitiello, G., (1986) *Nucl. Phys. B* **275**, 185-195.

45. Del Giudice, E., Preparata, G., & Vitiello, G. (1988) *Phys. Rev. Lett.* **61**, 1085-1185.

46. Bernardot F. (1992) *Europhysics Lett.* **17**, 34-44.

47. Tegmark, M. (2000) *Phys. Rev. E* **61**, 4194-4200.

48. Pereira, A., Kimble, A. & Peng A., (1992) *Phys. Rev. Lett.* **68**, 3663-3666.

49. Togerson, T., Branning, S., Monken, M. & Mandel A., (1995) *Phys. Lett. A* **204**, 323-328.

50. Sackett, L. (2000) *Nature* **404**, 256-259.

51. Julsgaard B., Kozhekin, A. Polzig, E. "Experimental long-lived entanglement of two macroscopic objects" *Nature* **413:** 400-403 (2001)

52. Rauschenbeutel E., *Science* **288,** 2024-2028 (2000)

53. Kwiat, P., Matle, P., Weifurter, P., & Zeilinger K., (1995) *Phys. Rev. Lett.* **75**, 4337-4343.

54. Altewischer, E., van Exter, M.P. & Woerdman, M.P. "Plasmon-assisted transmission of entangled photons",*Nature* **418** 304-306 (2002)

55. Oberparleiter, Weinfurter "High-efficiency entangled photon pair collection in type-II parametric fluorescence" *Phys. Rev. A***, 64***, 023802 (2001)*

56. Wolf, S.A., Awschalom, D. D., Buhrman, R. A. , Daughton, J. M. , von Molnár, S. , Roukes, M. L. , Chtchelkanova, A. Y. , & Treger D. M. "A Spin-Based Electronics Vision for the Future" *Science* **294** 1488-1495 (2001)

57. Zhang. S., *Nature Biotechnology* **21**, 10-18 (2003)

58. Pereira, A., Kimble, A. & Peng A. *Phys. Rev. Lett.* **68**, 3663-3666 (1992).